\documentstyle[epsf,epic,eepic,twoside]{book}
\def\thesection{\arabic{section}}
\newcommand{\sect}[1]{\stepcounter{section}\section*{\protect{
\noindent\large\bf\thesection~\normalsize\bf #1}}}

\def\thefigure{\arabic{figure}}
\def\fnum@figure{{\bf Figure \thefigure}}
\begin{document}
\def\pni{\par\noindent}

\pagestyle{myheadings}
\markboth{\qquad G.M. D'Ariano and M.F. Sacchi\hspace*{\fill}}
{\hspace*{\fill}
Repeatable two-mode phase measurement\qquad}

\thispagestyle{empty}
\vspace*{1mm}
\begin{center}
{\Large\bf Repeatable two-mode phase measurement}
\end{center}
\begin{center}
by {\large{\sl G.M. D'Ariano$^{a,b}$ and M. F. Sacchi$^{a,c}$}}, 
\end{center}
\begin{center}
$^a$ Dipartimento di Fisica \lq Alessandro Volta\rq,\\ 
Universit\`a degli Studi di Pavia, via A. Bassi 6, I-27100 Pavia, Italy\\
$^b$ Department of Electrical Engineering and Computer Science,\\
Northwestern University, Evanston, IL 60208, USA\\
$^c$ Istituto Nazionale di Fisica della Materia, Sezione di Pavia, Italy.\\ 
\end{center}

{
\begin{quote}{\bf Abstract.}
A measurement scheme to perform a repeatable phase detection on 
a two-mode field is presented. The interaction with the probe state 
(the output state of a phase-insensitive high-gain amplifier) is 
described by a Hamiltonian which is physically 
realizable in the rotating wave approximation. 
Information on the system is obtained through 
unconventional heterodyne measurements performed 
on the probe field after the interaction with the system. 
The expressions for the probability distribution and the state reduction 
are given.
\end{quote}}
\sect{Introduction}
The estimation of a phase shift on a two-mode field is not subjected 
to the limitations due to the complementarity principle that are 
suffered by a single mode field \cite{shsh,ban,shap}. 
A two-mode field corresponds to a complex photocurrent $\hat Z$, 
and a proper self-adjoint operator 
$\hat{\phi}=\arg(\hat{Z})$ is well-defined as long as 
$[\hat Z,\hat Z^{\dag}]=0$ \cite{shwa}. 
Indeed, the output photocurrent $\hat Z$ of an ideal heterodyne detector 
has just this property \cite{yuen}. For this reason, 
unconventional field heterodyning (i.e. with the 
signal and image-band modes both nonvacuum) has revealed promising 
possibilities \cite{shwa,rc} for ax exact phase measurement in terms of 
two-mode fields. In particular, after finding the physical 
(realizable) states that approach heterodyne eigenstates, we have 
shown \cite{rc} that the ideal sensitivity limit 
$\delta \phi =1/\overline{n}$ can be achieved for large mean number 
of photons $\overline{n}$. 
\pni In this paper, we provide a feasible scheme to perform a 
repeatable phase measurement on a generic two-mode field. 
The repeatability of the measurement is based on the fact 
that a heterodyne detection is carried out on the probe two-mode 
field only, after a suitable interaction with the system signal. 
The property of repeatability allows to check the evolution of 
the signal under successive measurements, and possibly 
also to drive the evolution 
itself by state reduction (i.e. selecting the state after measurement). 
Finally---more interesting for foundations---a 
repeatable phase measurement is a good candidate for detecting 
Schr\"odinger-cat states (see, for example, Ref. \cite{cat}).
\sect{Approaching the heterodyne eigenstates}
In Ref. \cite{rc} we presented a feasible scheme that achieves ideal phase 
detection on a two-mode field. The main idea 
is to perform an unconventional heterodyne measurement with non vacuum
image-band mode on a two-mode field, in order to get a probability 
distribution of the output photocurrent as sharp as possible. The 
signal photons are tuned to optimize the r.m.s. phase sensitivity
to the ideal limit $\delta \phi =1/\overline{n}$. 
\par As proved in Ref. \cite{yuen}, the output 
photocurrent of an ideal heterodyne detector corresponds to the complex 
operator $\hat{Z}=a+b^{\dag}$, where $a$ and $b$ denote 
the signal and the image-band mode, respectively. 
Indeed the heterodyne detector jointly measures the real and the 
imaginary part of $\hat{Z}$, which are expressed as a function of the 
quadratures $\hat c_{\phi}=\frac{1}{2}(c^{\dag}e^{i\phi}+\hbox{h.c.})$ of 
the single modes $c=a,b$ as follows
\begin{eqnarray}
\hat Z_1=\hbox{Re}\hat {Z}=\hat a_0+\hat b_0\qquad\qquad\hat Z_2=
\hbox{Im}\hat Z=\hat a_{\pi/2}-\hat b_{\pi/2}\;.\label{z12}
\end{eqnarray}
$\hat Z_1$ e $\hat Z_2$ are self-adjoint commuting operators, 
so that they can be jointly measured without 
the additional 3dB noise suffered by joint measurement of 
conjugated quadratures.\\
The eigenvectors of $\hat Z$ with complex eigenvalue $z=z_1+iz_2$
are given by the following two equivalent expressions \cite{shwa}
\begin{eqnarray}
|z\rangle\!\rangle &=& \int_{-\infty}^{+\infty}\frac{dx}{\sqrt{\pi}}
e^{2ixz_2}|x\rangle_{0}\otimes|z_1-x
\rangle_{0}\nonumber  \\
&=&\int_{-\infty}^{+\infty}\frac{dy}{\sqrt{\pi}}
e^{-2iyz_1}|y+z_2\rangle_{\pi /2}
\otimes |y\rangle_{\pi /2}\;.\label{zeta}
\end{eqnarray}
In Eq. (\ref{zeta}) the tensor product 
$|\psi'\rangle\otimes|\psi''\rangle$ denotes kets in the Hilbert space 
${\cal H}_a\otimes{\cal H}_b$ pertaining the two modes $a$ and $b$, 
whereas $|x\rangle_{\phi}$ denotes a quadrature 
eigenvector for phase $\phi$. 
The notation $|\ \rangle\!\rangle$ stands for two-mode states. 
For a number state representation of Eq. (\ref{zeta}), see Ref.~\cite{rc}. 
Of course, the sharpest density probability of the output 
photocurrent is obtained for a field in a state 
$|z\rangle\!\rangle$. 
The states $\{|z\rangle\!\rangle\}$ are Dirac-normalized 
and have infinite total number of photons, hence they are not 
physically realizable. Nevertheless, we showed in Ref. 
\cite{rc} that the ``twin-beams'' 
\begin{eqnarray}
|0\rangle\!\rangle_{\lambda}=(1-\lambda ^{2})^{1/2}  
\sum_{n=0}^{\infty}(-\lambda)^{n}|n\rangle\otimes|n\rangle
\;\label{zero}
\end{eqnarray}
at the output of a phase-insensitive 
amplifier (PIA) with gain $G=(1-\lambda ^{2})^{-1}$ 
approach the eigenstate $|0\rangle\!\rangle$ of $\hat Z$ 
(corresponding to zero eigenvalue) in the limit of infinite gain 
($\lambda\rightarrow 1^{-}$). 
The eigenstate $|z\rangle\!\rangle$ for nonzero eigenvalue is 
approximated by the physical state 
\begin{eqnarray}
|z\rangle\!\rangle_{\lambda}=e^{za^{\dag}-\overline{z}a}
(1-\lambda ^{2})^{1/2}  
\sum_{n=0}^{\infty}(-\lambda)^{n}|n\rangle\otimes|n\rangle
\;,\label{zetal}
\end{eqnarray}
which is achieved by combining the twin-beams 
$|0\rangle\!\rangle_\lambda $ with a strong coherent local 
oscillator at the frequency of the signal mode $a$ 
in a high-transmissivity beam splitter. 
The desired probability 
density of getting the value $z$ for the output photocurrent $\hat Z$ 
with the field in the state $|w\rangle\!\rangle_{\lambda}$ is given by
\begin{eqnarray}
|\langle\!\langle z|w\rangle\!\rangle_{\lambda}|^{2}  
=\frac{1}{\pi\Delta_{\lambda}^{2}}\exp\left( {-\frac{|z-w|^{2}}
{\Delta_{\lambda}^{2}}}\right)    \;,\label{prob}
\end{eqnarray}
where
\begin{eqnarray}
\Delta_{\lambda}^{2}=\frac{1-\lambda }{1+\lambda }\;.\label{D}
\end{eqnarray}
The state of the field $|w\rangle\!\rangle_{\lambda}$ 
has average number of photons
\begin{eqnarray}
\overline{n}={}_{\lambda}\langle\!\langle w|a^{\dag}a+b^{\dag}b
|w\rangle\!\rangle {}_{\lambda}=|w|^{2}+\frac{2\lambda^2}{1-\lambda^2}
\;\label{num}
\end{eqnarray}
and the marginal probability density
\begin{eqnarray}
p(\phi)=\int_{0}^{+\infty}\!d|z|
\,|z|\,|\langle\!\langle z|w\rangle\!\rangle_{\lambda}|^{2}  
\;
\end{eqnarray}
for the phase $\hat{\Phi}=\arg(\hat{Z})$ approaches for 
$\Delta_{\lambda}\ll |w|$ the Gaussian form
\begin{eqnarray}
p(\phi)\simeq\frac{|w|}{\sqrt{\pi}\Delta_{\lambda}}\exp\left[ -{|w|^2
\over\Delta_{\lambda}^2}(\phi -\theta)^2\right] \;,
\end{eqnarray}
where $\theta =\arg(w)$. The corresponding r.m.s. phase sensitivity 
$\delta\phi=\langle\Delta\phi^{2}\rangle^{1/2}$ is optimized 
in the limit of infinite gain at the PIA, achieving the ideal limit
\begin{eqnarray}
\delta\phi\simeq{1\over\bar n}\;\label{1sun}
\end{eqnarray}
for $|w|^2=(1-\lambda)^{-1}=\bar n/2$.
\pni The effect of nonunit quantum efficiency $\eta < 1$ of the 
heterodyne detector can be easily taken into account by changing 
Eq.~(\ref{D}) to
\begin{eqnarray}
\Delta_{\lambda}^2 \rightarrow \Delta_{\lambda}^2(\eta)=
\Delta_{\lambda}^2+{1-\eta\over\eta}\;,\label{Deta}
\end{eqnarray}
and the result (\ref{1sun}) still holds 
for $1-\eta\ll 2/{\bar n}$, otherwise sensitivity rapidly degrades 
towards shot noise.
\sect{Scheme for repeatable measurements}
On the line of the main results for the two-mode phase heterodyne 
detection summarized in the previous section, here we present a 
scheme for a repeatable phase measurement, giving also some hints for 
its experimental realization. Our approach allows to perform a phase 
measurement on a generic two-mode field, without destroying it: 
in the following we compute the reduced state, 
depending on the outcome of the measurement.
\pni We propose an interaction Hamiltonian bilinear in the four 
field modes $a,b$ (for the system) and $c,d$ (for the probe) as follows
\begin{eqnarray}
\hat H = -K{i\over 2}\left[(a^{\dag}c+bc+ad+b^{\dag}d)-
\hbox{h.c.}\right]\;,
\label{h2m}\end{eqnarray}
where $K$ is a coupling constant.\\
From definitions (\ref{z12}), and 
introducing the complex current for the probe field
\begin{eqnarray}
\hat A\equiv c+d^{\dag}=\hat A_1+i\hat A_2\;,\label{apr}
\end{eqnarray}
the Hamiltonian (\ref{h2m}) rewrites
\begin{eqnarray}
\hat H=K\left[\hat Z_1(\hat c_{\pi/2}+\hat d_{\pi/2})-\hat Z_2(\hat c_0-
\hat d_0)\right].\;
\end{eqnarray}
The Heisenberg evolution of a probe operator of the form $f(\hat A)$ 
for an interaction time $\tau=\hbar/K$ is
\begin{eqnarray}
\hat U^{\dag}\,f(\hat A)\,\hat U
=f(\hat A+\hat Z)
\;,\label{heis}
\end{eqnarray}
where 
\begin{eqnarray}
\hat U=\exp(-i\hat H \tau)=\exp\left\{ -i\left[\hat Z_1(c_{\pi/2}+d_{\pi/2})-
\hat Z_2(c_0-d_0) \right] \right\} \;\label{unitt}
\end{eqnarray}
is the unitary evolution operator.\\
After the interaction, the probe modes $c$ and $d$ are heterodyne 
measured, with $c$ as the signal mode and $d$ as the image-band mode. 
As explained in the previous section, this corresponds 
to measure the photocurrent $\hat A$. 
Notice that here we no longer need a beam splitter for 
displacing the state, because the interaction Hamiltonian itself 
transfers signal to the twin-beams before heterodyne 
detection. This indirect measurement 
provides information about the probability density of the complex 
eigenvalue $z$ pertaining the system operator $\hat Z$. 
The probability density is computed through the relation
\begin{eqnarray}
\hbox{Tr}_S[\hat F(z)\hat \rho_S]=\hbox{Tr}_{S,P}[|z\rangle\!\rangle
\langle\!\langle z|
\,\hat U\,(\hat \rho _S\otimes\hat\rho_P)\,\hat U^{\dag}]\;,\label{zz}
\end{eqnarray}               
where $\hat F(z)$ is a probability operator-valued measure (POM) and 
$|z\rangle\!\rangle\langle\!\langle z|$ represents an orthogonal 
projector on the eigenspace of the two-mode probe Hilbert space 
${\cal H}_c\otimes{\cal H}_d$ 
relative to the eigenvalue $z$ of $\hat A$. 
The tensor product 
$\hat \rho _S\otimes\hat\rho _P$ denotes the (disentangled) state of 
system and probe before the interaction. 
For a probe preparation in the twin-beams state 
(\ref{zero}), the POM $\hat F(z)$ writes
\begin{eqnarray}
\hat F(z)={}_{\lambda }\langle\!\langle 0|\hat U^{\dag}|z\rangle\!
\rangle\langle\!\langle z|\hat U|0\rangle\!\rangle_{\lambda }\doteq
\hat\Omega ^{\dag}(z)\,\hat \Omega(z)
\;.\label{fz0}\end{eqnarray}               
From the Eqs. (\ref{zeta}), (\ref{unitt}) and (\ref{prob}) 
with $w\!=\!0$, taking into account the additional phase 
${}_{\lambda }\langle\!\langle 0
|z\rangle\!\rangle =|{}_{\lambda }\langle\!\langle 0
|z\rangle\!\rangle |e^{iz_1z_2}$ \cite{nota}, 
the system ``amplitude-operator'' $\hat\Omega ^{\dag}(z)$ 
is evaluated as follows
\begin{eqnarray}
\hat\Omega ^{\dag}(z)&=&
\exp\left\{  i\hat Z_1\left( i{\partial\over\partial
z_1}+z_2\right)+i\hat Z_2\left(i{\partial\over\partial
z_2}+z_1 \right)\right\}{}_{\lambda }\langle\!\langle 0
|z\rangle\!\rangle \nonumber \\
&=&{e^{iz_1z_2}\over \sqrt{\pi}\Delta_{\lambda}}\,\exp\left( -{
\left|\hat Z-z\right|^2\over 2\Delta_{\lambda}^2}\right) \;,\label{omg}
\end{eqnarray}
where $\Delta _{\lambda} $ is given by Eq. (\ref{D}). As a consequence, 
one gets the expression for the POM 
\begin{eqnarray}
\hat F(z)={1\over \pi\Delta_{\lambda}^2}\,\exp\left( -{
\left|\hat Z-z\right|^2\over \Delta_{\lambda}^2}\right) \;,\label{fz}
\end{eqnarray}
which provides the probability density
\begin{eqnarray}
P(z\|\hat\rho _S)=\hbox{Tr}[\hat F(z)\hat\rho_S]={1\over\pi\Delta_{\lambda}^2}
\int_C d^2z' \,{}_S\langle\!\langle z'|\hat \rho_S|z'\rangle
\!\rangle _S\,\exp\left(-{\left|z'-z\right|^2
\over \Delta_{\lambda}^2}\right)
\;.\label{conv}
\end{eqnarray}
Here $\{|z'\rangle\!\rangle _S\}$ are the eigenstates of the system 
operator $\hat Z$ and the integral is over the complex plane. 
Eq.~(\ref{conv}) is a convolution of the ideal probability 
with a Gaussian that narrows for increasing gain of the PIA. 
Notice that the result (\ref{fz}) can be derived more easily 
through Eqs.~(\ref{fz0}) and (\ref{heis}), upon 
defining formally the projector for $\hat A$ as a Dirac delta on the 
complex plane, namely
\begin{eqnarray}
|z\rangle\!\rangle\langle\!\langle z|=\delta^{(2)}(\hat A -z)\;,
\end{eqnarray}
then obtaining from Eq. (\ref{fz0}):
\begin{eqnarray}
\hat F(z)={}_{\lambda }\langle\!\langle 0|
\delta^{(2)}(\hat A -\hat Z-z)|0\rangle\!\rangle_{\lambda }
\;;
\end{eqnarray}
hence, using Eq. (\ref{prob}), one gets the result.\\
The operator $\hat\Omega (z)$ has been explicitly computed because its 
adjoint action on the system state $\hat\rho_ S$ 
provides the reduced state  $\hat\rho _z$ 
after the measurement with outcome $z$. One has
\begin{eqnarray}
\hat\rho _z={\hat\Omega(z)\,\hat \rho _S\,\hat\Omega ^{\dag}(z)\over 
\hbox{Tr}[\hat F(z)\hat\rho _S]}=
{1\over \pi\Delta_{\lambda}^2}\,{\exp\left(-{
\left|\hat Z-z\right|^2\over 2\Delta_{\lambda}^2}\right)\,\hat\rho _S
\,\exp\left( -{\left|\hat Z-z\right|^2\over2 \Delta_{\lambda}^2}\right )
\over \hbox{Tr}[\hat F(z)\hat\rho _S]}  \;.\label{rhoz}
\end{eqnarray}
The POM that provides the probability density for the phase is 
the marginal one of $\hat F(z)$ in Eq.~(\ref{fz}), namely
\begin{eqnarray}
d\hat\mu(\phi)&=&\int_{0}^{+\infty}d|z|\,|z|\hat F(z)\nonumber \\
&=&{1\over 2\pi}\exp\left( -{|\hat
Z|^2\over\Delta_{\lambda}^2}\right)+{1\over\pi\Delta_{\lambda}}\hbox{Re}\left(
\hat Ze^{-i\phi}\right)\exp\left\{
-{1\over\Delta_{\lambda}^2}\left[\hbox{Im}\left(\hat Ze^{-i\phi}\right) 
\right]^2\right\}
\nonumber \\&\times&\,{\sqrt{\pi}\over2}\,\left\{1+\hbox{erf}
\left[ {\hbox{Re}\left(\hat Ze^{-i\phi}\right) \over\Delta_{\lambda}}\right]
\right\}   \;,\label{muph}
\end{eqnarray}
with $\phi =\arg(z)$. 
In the limit of infinite average number of photons at the twin-beams 
($\Delta_{\lambda}\rightarrow 0$), Eq.~(\ref{muph}) 
approaches the ideal POM
\begin{eqnarray}
d\hat\mu(\phi)=\delta\left(\hat\Phi -\phi\right) \;
\end{eqnarray}
where $\hat \Phi=\arg (\hat Z)$.
\pni As regards the case of a heterodyne detection with quantum 
efficiency $\eta < 1$, the projector in the right side of Eq.(\ref{zz})
needs to be replaced by the POM \cite{bilkent}
\begin{eqnarray}
\hat A_{\eta}(z)={\eta\over \pi (1-\eta)}\,\exp\left( -{\eta\over 1-\eta}
\,|\hat A -z|^2\right) \;.\label{eta2}
\end{eqnarray}
This increases the variance (\ref{Deta}) of the POM's 
(\ref{fz}) and (\ref{muph}). The corresponding reduced state 
$\hat \rho_{z}^{(\eta)}$ becomes
\begin{eqnarray}
\hat \rho_{z}^{(\eta)}=\int _{C}d^2z'\ 
{\eta\over \pi (1-\eta)}\,e^{-{\eta\over 1-\eta}
\,|z'-z|^2}\hat\Omega(z')\,\hat \rho _S\,\hat\Omega ^{\dag}(z')
\,\hbox{Tr}[\hat F_{\eta}(z)\hat\rho _S]^{-1}\;,\label{rhozeta}
\end{eqnarray}
where $\hat\Omega(z')$ is the same as in Eq. (\ref{omg}) and 
$\hat F_{\eta}(z)$ is the new POM after the substitution 
(\ref{Deta}).\\
Eq. (\ref{rhozeta}) displays a conceptually noteworthy difference from 
Eq.~(\ref{rhoz}): the ideal measurement reduces a pure initial 
state into a pure state (it is a quasi-complete measurement, according 
to Ozawa's definition \cite{ozw}), 
whereas a nonunit quantum efficiency leads to mixing.
\pni Regarding the experimental feasibility of the 
repeatable measurement scheme 
here presented, we notice that 
Hamiltonian (\ref{h2m}) can be achieved 
in the parametric approximation by means of two classical 
undepleted pumps. The interaction Hamiltonian in the Dirac picture 
is obtained in the rotating wave approximation from a 
non linear susceptibility $\chi ^{(2)}$ ({\em three-wave mixing}). 
Indeed, the following frequency arrangement of the probe mode 
$d$ and the pump modes $\gamma ,\xi$ in comparison with the probe mode 
$c$ and the system modes $a,b$ ($\omega_a<\omega_b$)
\begin{eqnarray}
\left\{\begin{array}{l}
\omega _d=\omega_c+\omega _b-\omega _a\\
\omega _{\xi}=\omega _c-\omega _a\\
\omega _{\gamma}=\omega _c+\omega _b 
\end{array}\right. 
\;\end{eqnarray}
with the restrictions
\begin{eqnarray}
&&\omega_b\neq2\omega_a\,;\qquad\qquad\omega_c>\omega_b \nonumber \\
&&\omega_c\neq {3\over 2}\omega_a\,,2\omega_a\,,
\omega_a+{\omega_b\over 2}\,,\omega_a+\omega_b\,,2\omega_a+\omega_b
\;
\end{eqnarray}
insure that the only surviving terms 
in the rotating wave approximation are represented by the Hamiltonian
\begin{eqnarray}
\hat H \propto\left[\left(a^{\dag}c\xi^{\dag}+b^{\dag}d\xi^{\dag}+
ad\gamma^{\dag}+bc\gamma^
{\dag}\right) +\hbox{h.c.}\right]\;.\label{hex}
\end{eqnarray}
The Hamiltonian (\ref{hex}) clearly coincides 
with the Hamiltonian in Eq. (\ref{h2m}) in the parametric 
approximation of undepleted pumps. 
It is clear that for a 
suitable frequency arrangement, one could also use 
a {\em four-wave mixing} $\chi^{(3)}$ medium.
\section*{\normalsize\bf References}
\begin{description}
\bibitem[1]{shsh} J. H. Shapiro and S. R. Shepard, Phys. Rev. A {\bf 43}, 
3795 (1991).
\bibitem[2]{ban} M. Ban, Phys. Rev. A {\bf 50}, 2785 (1994). 
\bibitem[3]{shap} J. H. Shapiro, Physica Scripta T {\bf 48}, 105 (1993). 
\bibitem[4]{shwa} J. H. Shapiro and S. S. Wagner, IEEE J. Quantum Electron. 
QE {\bf 20}, 803 (1984).
\bibitem[5]{yuen} H. P. Yuen and J. H. Shapiro, 
IEEE Trans. Inform. Theory IT {\bf 26}, 78 (1980).
\bibitem[6]{rc} G. M. D'Ariano and M. F. Sacchi, Phys. Rev. A {\bf 52}, 
R4309 (1995)
\bibitem[7]{cat} G. M. D'Ariano, M. Fortunato and P. Tombesi, Nuovo 
Cimento B {\bf 110}, 1127 (1995).
\bibitem[8]{nota} This phase relation can be derived from the number 
representation of vectors $\{|z\rangle\!\rangle\}$ given in Ref. \cite{rc}.
\bibitem[9]{ozw} M. Ozawa, J. Math. Phys. {\bf 27}, 759 (1986).
\bibitem[10]{bilkent} G. M. D'Ariano, {\em Concepts and Advances in 
Quantum Optics and Spectroscopy of Solids}, Eds. T.Hakio\v{g}lu and 
A.S. Shumovsky (Bilkent Univ.), Kluwer Academic Publishers 1996 (in press).
\end{description}
\end{document}